\documentclass[journal]{IEEEtran}
\usepackage[tight,footnotesize]{subfigure}
\usepackage{enumitem}
\usepackage{float}
\usepackage{multicol,multienum}
\usepackage{breqn}
\usepackage{stfloats}
\usepackage{multirow}
\usepackage{diagbox}
\usepackage{hyperref}
\usepackage{bm}
\usepackage{cite}
\usepackage{graphicx}
\usepackage{tabularx}
\usepackage{algorithm}
\usepackage{algorithmic}
\usepackage{booktabs}

\usepackage[tight,footnotesize]{subfigure}
\usepackage{url}
\usepackage{doi}

\ifCLASSINFOpdf

\else
 
\fi

\begin{document}
\title{A Secure Federated Learning Framework for 5G Networks}
\author{Yi~Liu,
        Jialiang Peng,
       Jiawen Kang,
       Abdullah M. Iliyasu,~\IEEEmembership{Member,~IEEE,}
       Dusit Niyato,~\IEEEmembership{Fellow,~IEEE,}
       and Ahmed A. Abd El-Latif
\thanks{Yi Liu, and Jialiang Peng are with the School of Data Science
and Technology, Heilongjiang University, Harbin 150080, China  (e-mail: 97liuyi@ieee.org; jialiangpeng@hlju.edu.cn).}
\thanks{Jiawen Kang is with Energy Research Institute, Nanyang Technological University, Singapore (e-mail: kavinkang@ntu.edu.sg).}
\thanks{Dusit Niyato is with School of Computer Science and Engineering, Nanyang Technological University, Singapore (e-mail: dniyato@ntu.edu.sg).}
\thanks{Abdullah M Iliyasu is with Electrical Engineering Department, College of Engineering, Prince Sattam Bin Abdulaziz University, Al-Kharj 11942, Saudi Arabia and with School of Computing, Tokyo Institute of Technology, Yokohama 226-8502, Japan. Also, he with School of Computer Science and Technology, Changchun University of Science and Technology, Changchun 130022, China (e-mail:a.iliyasu@psau.edu.sa; abdul-m-elias@ieee.org)}
\thanks{Ahmed A. Abd El-Latif is with the Mathematics and Computer Science Department, Faculty of Science,
Menoufia University, Shebin El-Koom 32511, Egypt, and is with School of Information Technology and Computer Science, Nile University, Egypt (e-mail: a.rahiem@gmail.com; aabdellatif@nu.edu.eg).}
\thanks{Jialiang Peng is the corresponding author.}}

\markboth{IEEE Wireless Communications Magazine }
{Shell \MakeLowercase{\textit{et al.}}: Bare Demo of IEEEtran.cls for IEEE Journals}

\maketitle

\begin{abstract}
Federated Learning (FL) has been recently proposed {as an emerging paradigm to build machine learning models using} distributed training datasets that are locally stored and maintained on different devices in 5G networks while providing privacy preservation for participants. In FL, the central aggregator accumulates local updates uploaded by participants to update a global model. However, there are two critical security threats: poisoning and membership inference attacks. These attacks may be carried out by malicious or unreliable participants, resulting in the construction failure of global models or privacy leakage of FL models. Therefore, it is crucial for FL to develop security means of defense. In this article, we propose a blockchain-based secure FL framework to create smart contracts {and} prevent malicious or unreliable participants from involving in FL. {In doing so}, the central aggregator recognizes malicious and unreliable participants by automatically executing smart contracts to defend against poisoning attacks. {Further}, we use local differential privacy techniques to prevent membership inference attacks. Numerical results  {suggest} that the proposed framework can effectively {deter} poisoning and membership inference attacks, thereby improving the security of FL in 5G networks.
\end{abstract}


\IEEEpeerreviewmaketitle
\section{Introduction}
{Distributed machine learning has spawned a lot of useful applications, such as on-device learning and edge computing \cite{ref-4}. However, due to communication delays and network bandwidth limitations in fourth-generation (4G) networks, mobile users holding smart devices with limited computing power cannot fully participate in distributed machine learning tasks. Fortunately, in the fifth-generation (5G) networks, {bottlenecks attributed to communication latency and network bandwidth will be overcome} \cite{ref-3}. {Therefore, attention can be focused on addressing} the performance and efficiency issues of distributed machine learning can be addressed. {Consequently,} mobile devices will be able to participate in distributed machine learning.} However, traditional distributed machine learning techniques require a certain amount of private data to be aggregated and analyzed at central servers during the model training phase \cite{ref-7}. Such a training process would lead to potential privacy leakage for users in 5G networks \cite{ref-1,ref-6}.

To address such privacy challenges, McMahan et al. \cite{ref-11} proposed a collaboratively distributed machine learning paradigm for mobile devices called Federated Learning (FL). {FL involves the training machine learning models over devices \footnote{We use the term `device' throughout this article to describe entities in 5G networks, such as edge nodes, clients, smartphones, etc.} or siloed data centers while {maintain}ing the training datasets locally without sharing raw training data \cite{ref-12}.} The procedure of FL {execution} is divided into three phases: {initialization, aggregation, and update phases}. In the initialization phase, the central aggregator presents a pre-trained global model on a public dataset (e.g., MNIST\footnote{\href{http://yann.lecun.com/exdb/mnist/}{http://yann.lecun.com/exdb/mnist/}}, CIFAR-10\footnote{\href{http://www.cs.toronto.edu/~kriz/cifar.html}{http://www.cs.toronto.edu/~kriz/cifar.html}}), to each device. {Following that, each device uses 5G networks to train and improve the current global model on the local dataset in each iteration.}
In the aggregation phase, the central aggregator aggregates local model updates (i.e., gradient information) from the devices. In the update phase, the central aggregator aggregates all local model updates to generate a new global model for the next iteration. Both the devices and the central aggregator repeats the above process until the global model reaches a certain accuracy or optimal convergence. This paradigm significantly reduces the risks of privacy leakage by decoupling the model training from direct access to the raw training data \cite{ref-12}.

{Although FL brings new application scenarios (e.g., edge computing and on-device learning) to devices in 5G networks, it still suffers from two critical threats: poisoning and membership inference attacks \cite{ref-1-1}.}
Firstly, if  {the poisoning attack} occurs, an unreliable device submits an ``error update'', FL model would be ``poisoned.'' \cite{ref-10} A vicious attack will cause a failure in FL global model update. Secondly, in the case of the {membership inference} attack, since the intermediate gradients in training models always contain rich semantic information (e.g., model parameters), {adversaries could exploit reverse engineering techniques to access} some sensitive information (e.g., raw distribution of training data) \cite{ref-13-13}.
{Therefore, these vulnerabilities compel the
need to develop a secure and privacy-preserving framework for FL, where the global model updates can be aggregated more securely.}

{In this study, we propose a secure FL framework that {exploits} the capabilities of smart contracts in blockchain to defend against poisoning attacks and also introduce local differential privacy techniques to {mitigate} inference attacks. Specifically, smart contracts allow a task publisher to reward participants who can {provide well-trained} local models for a specific FL task \cite{ref-14}. The proposed framework establishes a marketplace to train FL models in an automated and anonymous manner for each participant (i.e., mobile device) through 5G networks.} The local differential privacy technique provides robust defensive measures against membership inference attacks by adding well-designed noises to the model updates during the model training phase \cite{ref-13-13}. Based on the proposed framework, a central aggregator can create a decentralized and trustless marketplace to eliminate the impact of overreliance on participants in FL.


{In 5G communication systems, task publishers can be typically deployed on edge nodes with sufficient storage and data processing capabilities, and mobile devices can easily access edge nodes to train local models.}
The proposed framework creates a marketplace where participants that are skilled in solving FL problems can profit directly from smart contracts \cite{ref-14}. The main contributions of this study are summarized as follows:
\begin{itemize}
	\item We design a blockchain-based federated learning framework to achieve secure and reliable federated learning as well as all-round defense against poisoning attacks.
	\item We introduce the local differential privacy technology as a firewall against membership inference attacks on federated learning models.
	\item We create a marketplace where participants endowed in solving federated learning problems can benefit directly from their skills. This will provide a trustworthy platform to motivate participants to create better federated learning models.
\end{itemize}

\section{Federated Learning and Its Threats}
\subsection{Federated Learning and Its Application Scenarios in 5G Networks}
Federated Learning \cite{ref-11,ref-15} is a collaborative machine learning framework that required no centralized training data. {In FL, the local devices download the global model through 5G networks from the central aggregator, and then these devices train and improve the current global model by using their local raw data \cite{ref-11}.}
{Typically}, each device trains its local model using Distributed Stochastic Gradient Descent {(D-SGD)} algorithm \cite{ref-17} and uploads the model updates (i.e., gradient information) to a central aggregator \cite{ref-12}. The aggregator updates a new global model by collecting all local updates and calculating the mean value of these local model updates. Federated learning aims to build machine learning models based on datasets stored locally in distributed devices without compromising privacy.

Some typical FL application scenarios in 5G networks are listed as follows.
\begin{itemize}
	\item \textit{Internet of Things (IoT)}: ByteLAKE  and Lenovo  present  a technique called Federated Intelligence IoT\footnote{\href{https://becominghuman.ai/theres-a-better-way-of-doing-ai-in-the-iot-era-feabbbc1b589}{https://becominghuman.ai/theres-a-better-way-of-doing-ai-in-the-iot-era-feabbbc1b589}} that  not only enables IoT devices in 5G networks to learn from each other but also makes it possible to leverage local machine learning models on IoT devices. {In such scenarios, FL could be} applied to build a {personalized voice assistant} without compromising personal privacy.
	\item \textit{Internet of Vehicles (IoV)}: The Shenzhen Municipal Government in China cooperates with commercial organizations such as DiDi Chuxing and HelloBike companies to conduct the real-time traffic flow forecasting \cite{ref-4} based on IoV  in 5G networks. Meanwhile, General Data Protection Regulation (GDPR)\footnote{\href{https://eugdpr.org/}{https://eugdpr.org/}} prohibits any organization from directly trading personal data due to concerns about privacy leakage. Therefore, FL can be applied to build a traffic flow prediction model that not only improves the prediction accuracy but also protects personal privacy.
\end{itemize}


\subsection{{Threats and Countermeasures}}
Although FL has many promising applications in scenarios related to 5G networks, two threats have been identified as obstacles to its full-scale deployment. {Noting that it is assumed that the central aggregator is not compromised or malicious, we enunciate the two threats as follows:}
\begin{itemize}
	\item \textit{Threat 1: Participants with malicious behaviors.} Participants may have intentional or unintentional malicious behaviors during FL training phase \cite{ref-12}. Intentional malicious participants can submit wrong model updates, failing the update of the FL model. Unintentional malicious participants may upload model updates that have a negative effect on global model updates as they utilize low-quality training data. When the central aggregator aggregates these local model updates to update the global model, it eventually makes a poor accuracy or even useless global model.
All of these intentional or unintentional malicious behaviors {may ``poison'' the FL
model}. In a word, the current FL {models rely} on a trust mechanism, which {makes it} vulnerable to suffer from poisoning attacks \cite{ref-16}.
	\item \textit{Threat 2: Disclosure of sensitive information.} In FL, each participant uploads the updated parameters of the own local training model to the central aggregator. However, existing studies show that adversaries can still initiate a membership inference attack to obtain sensitive information from the updated parameters associated with the local data. Specially, adversaries can utilize the sensitive information disclosed by network snooping attacks in 5G networks to threaten the privacy of FL model.
\end{itemize}

{To address the poisoning attack, in \cite{ref-12}, Kang et al. {explored using blockchains to build} a decentralized market using identity and reputation systems to deter poisoning attacks.} {However, such a method favours the gradient direction for updates uploaded from participants with high reputation value, which leads to the poor generalization ability of FL.} {Moreover, for} local sensitive information leakage problems, existing studies show that the differential privacy technique is one of the best solutions \cite{ref-13-13}. {Therefore,  we introduce a local differential privacy technique to address the poisoning attack issues.
 {Guided by the above considerations, we outline the main objectives in the proposed FL framework.}}
\begin{itemize}
	\item {\textit{Establishing fairness trading market for participants in blockchain-based FL framework.}} {Based on Ethereum}, the proposed framework can create a market for collaborative training of FL models anonymously. Note that the proposed framework can be also applied to other blockchain platforms. In this marketplace, {honest participants can make profits by submitting  correct solutions to FL problems.} {To ensure the security of framework, the central aggregator verifies the updates of local models based on smart contracts.}
The need for mutual trust between participants in this framework is eliminated because the protocol uses cryptographic authentication to secure all transactions.

\item \textit{Protecting sensitive information in federated learning.}
{In the proposed framework, participants add well-designed noises to their model updates by applying the local differential privacy technique. It utilizes some noise-adding mechanisms (e.g., Gaussian Noise Mechanism or Laplace Noise Mechanism) to {secure model parameters.}} Therefore, even if adversaries {gain access to} the noise-adding gradient information, they cannot recover the original model parameters and local data.
\end{itemize}

\section{Secure Federated Learning Framework}
\begin{figure*}[!t]
	\centering
	\includegraphics[width=0.8\linewidth]{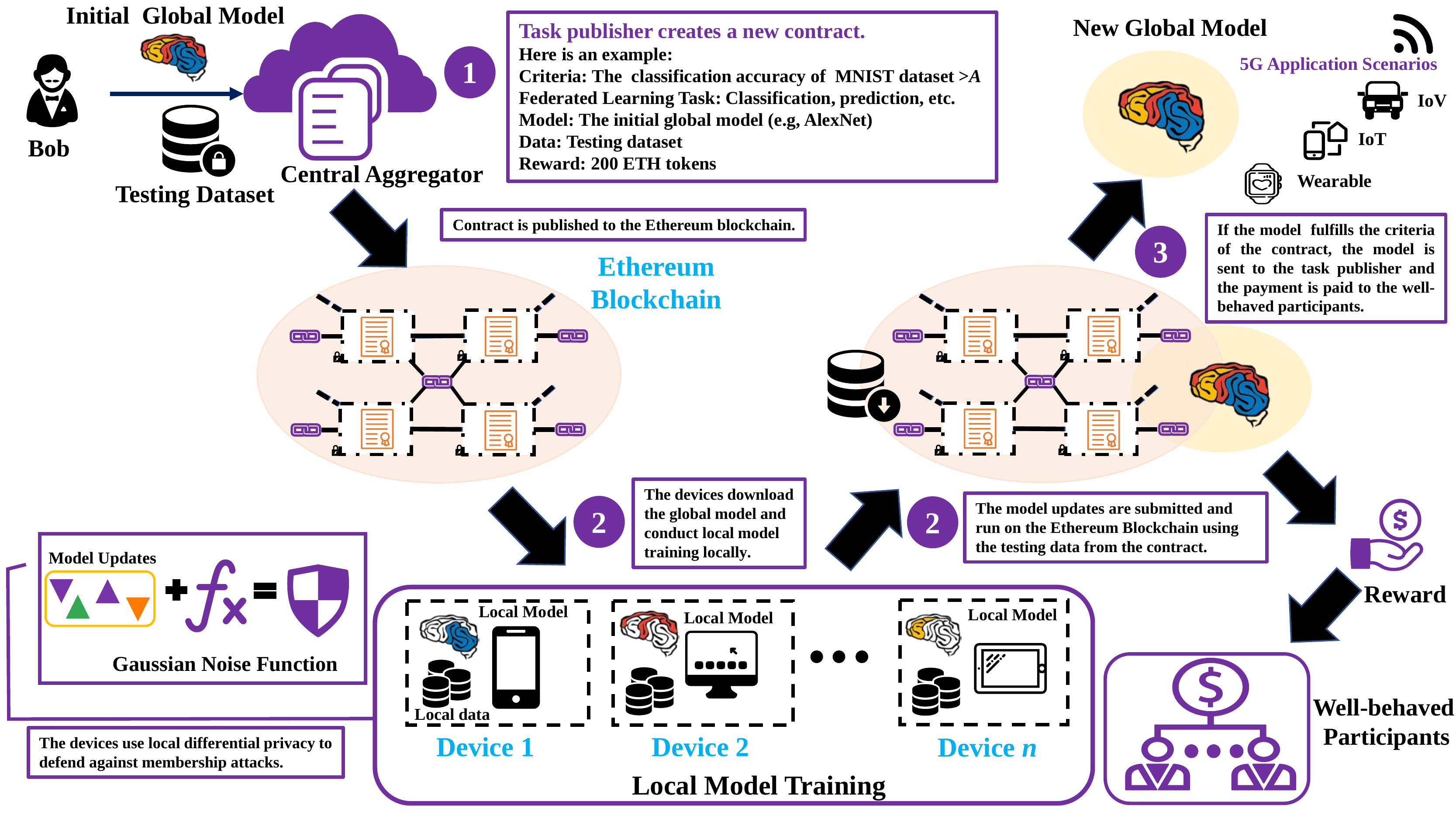}
	\caption{{An overview of proposed secure federated learning framework.}}
	\label{fig-1}
\end{figure*}

We first describe the related concepts and terms used in the proposed Secure Federated Learning (SFL) framework.
\begin{itemize}
	\item {\textit{Smart contracts:}} {Blockchain-based Ethereum implementation of a smart contract.}
	\item \textit{Wallet address:} an escrow account of Ethereum that {stores} the rewards.
	\item \textit{Task publisher:} anyone who can publish FL tasks and interacts with {the} central aggregator.
	\item \textit{Central aggregator:} an entity  responsible for aggregating model updates uploaded by devices and interacts with the task publishers.
	\item \textit{Device:} a physical entity in 5G networks involving in FL.
	\item  \textit{Local model:} the model that a device trains with its local data.
    \item \textit{Global model:} an initial global model that is issued by the task publisher for the devices, and global models are updated by the central aggregators during model iterations.
	\item \textit{Model update:} the gradient information generated by local model training.
	{\item \textit{Data points:} points made up of inputs and outputs.}
	{\item \textit{Data group:} a matrix made up of data points.}
\end{itemize}
{ Fig. \ref{fig-1} further illustrates the pipeline of the proposed SFL framework. In this framework, we introduce the smart contract technique to deter the poisoning attacks (see Fig. \ref{fig-2}), and  the local differential privacy technique is applied to curtail the inference attacks (see Fig. \ref{fig-3}).}
\begin{itemize}

\item \textit{Phrase 1, Initialization}. A task publisher named Bob creates a smart contract including a testing dataset, an initial global model, evaluation criteria, and a reward amount.
The accuracy performance is used as the evaluation criterion to evaluate the training quality of FL model. The reward is a monetary reward, such as ``ETH'' tokens. The above smart contract is published to the Ethereum blockchain. The central aggregator sends the initial global model to participants in Ethereum.
\item \textit{Phrase 2, Aggregation}. Each participant downloads the initial global model to train the local model  {using} its local dataset.
When a device succeeds in training a local model, it submits its local model updates to the Ethereum blockchain. {All the Ethereum blockchain miners use the evaluation function in the smart contracts to evaluate all uploaded model updates and generate an average value of model quality for each uploaded model. {We note that, similar to reference \cite{ref-11}, a federated averaging algorithm (i.e., average aggregation) is used in the proposed framework}. Participants use a differential privacy technique to add well-designed noise to the uploaded model updates. The high-quality (i.e., high accuracy) model updates with larger values (e.g., {if it is larger than a given threshold determined by task publishers)} will be sent to the central aggregator.}


\item \textit{Phrase 3, Update}. The task publisher gets the current global model from the central aggregator, and then  {prepares} for the next round of iterative training.

\end{itemize}

{The proposed SFL framework allows participants in the blockchain to implement local model training for rewards in a trustless manner.  {Moreover,} the participants will be rewarded by the efforts of their high accuracy model updates for FL tasks.} As presented in Fig. \ref{fig-2}, the detailed steps  {of} the phases mentioned above are described as follows.
\subsection{Initialization Phase}
\textbf{Step 1:}  The task publisher creates a smart contract with the following operations.
\begin{itemize}
	\item The initialization function \underline{init()} is called to initialize SFL  {framwork}.
	\item The number of iterations is  {based on} the size of the blocks.
	\item The task publisher hashes the testing dataset into the hashed data groups using a certain hash function.
\end{itemize}

\textbf{Step 2:} The task publisher hashes the indices of the data groups and sends this hashed result as the seed of random numbers to the central aggregator. The task publisher stores the reward in an encrypted address.  {Next}, the central aggregator calls the \underline{init()} function to initialize SFL framework.

\textbf{Step 3:} The central aggregator verifies the testing data with the hash value previously provided for  {the data groups and writes} the testing data groups and the random numbers into the smart contracts. The devices in 5G networks call the \underline{get\_global\_model()}  and \underline{get\_FL\_task()} functions to receive the global model and FL task sent by the central aggregator, respectively.

At this point, the smart contracts initialization in SFL framework has been completed. Devices can begin to train their local models for a particular FL task.

\subsection{Aggregation Phase}
\textbf{Step 1:} \textit{Local model training}.
In this step, each device  {calls} the \underline{local\_model\_training()} function to use its local data to train the local model. Without loss of generality, it is assumed that  D-SGD algorithm \cite{ref-17} is  {used} to update the local model updates.

\begin{figure*}[!t]
	\centering
	\includegraphics[width=0.9\linewidth]{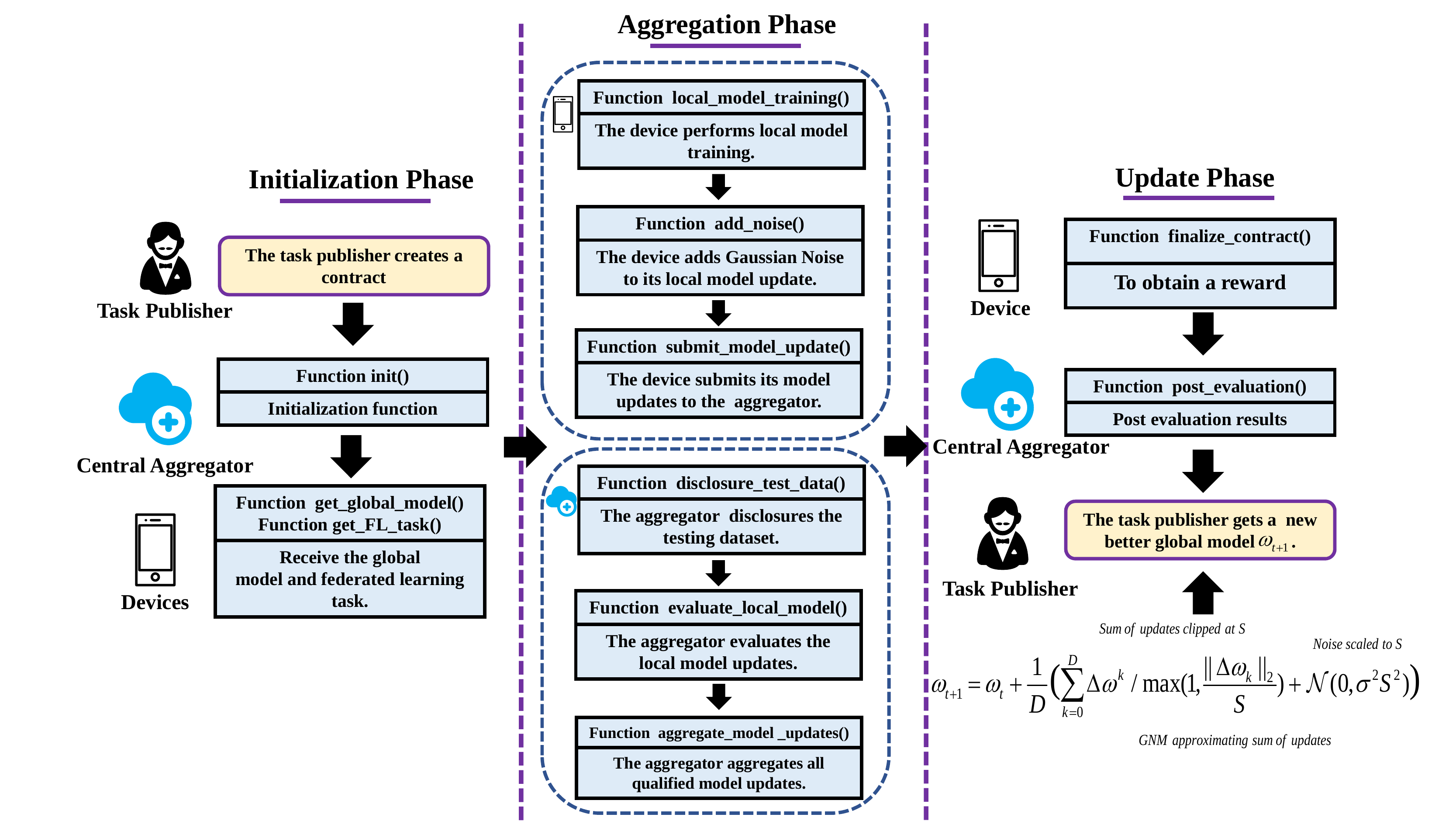}
	\caption{{The workflow in proposed secure federated learning framework.}}
	\label{fig-2}
\end{figure*}

\begin{figure}[!htb]
	\centering
	\includegraphics[width=0.9\linewidth]{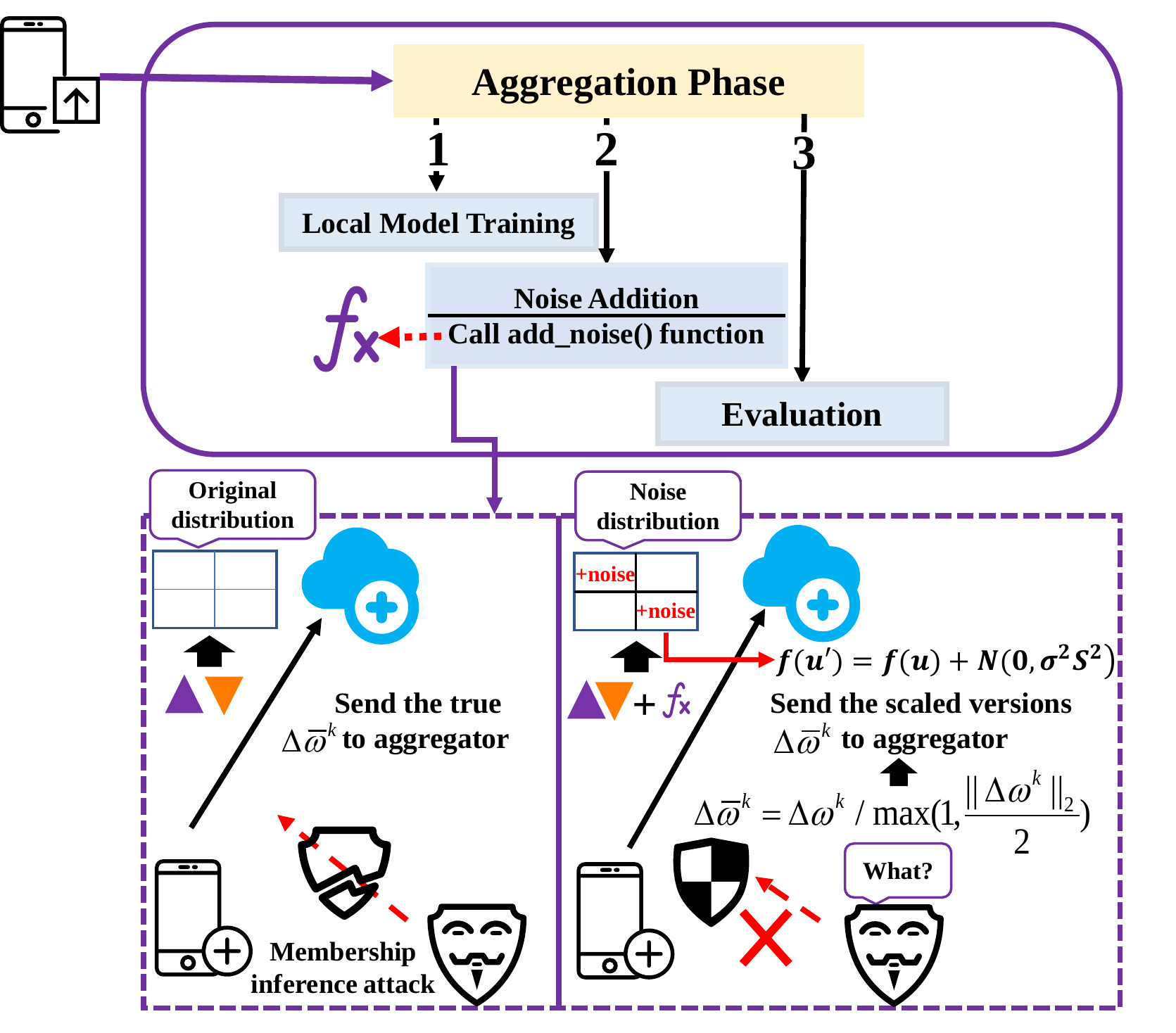}
	\caption{{Illustration of three steps of aggregation phase}: (1) Each device performs
local model training and prepares to submit its local model update, (2) Each device uses the differential privacy technique to add Gaussian Noise on the local model update, (3) Each  device
executes the \underline{submit\_model\_update()} function to submit the model update.}
	\label{fig-3}
\end{figure}

\textbf{Step 2:} \textit{ {Noise Addition.}} As shown in Fig. \ref{fig-3}, each device uses Gaussian
Noise Mechanism (GNM) to add noise on the local model updates  {prior to submission} to the central aggregator.
To defend against membership inference attacks, each device $d_i$ first perturbs its model updates $u_i$ using a randomized perturbation function $f$ based on GNM. {Next}, $d_i$ sends the perturbed model updates $u_i^'$ to the aggregator instead of its original model updates $u_i$.
To achieve this goal, we require that the function $f$ satisfies the requirement of $(\varepsilon ,\delta )$-local differential privacy ($(\varepsilon ,\delta )$-LDP) \cite{ref-13-13}, where $\varepsilon  > 0$ is a privacy parameter to make a tradeoff between the security and the availability, and $\delta$ is defined as the probability that plain $\varepsilon$-LDP is broken.
$f(u')$ is denoted as a fixed-scale Gaussian noise added to $f(u)$,  and the sensitivity $S$ is defined as the maximum  absolute distance between $f(u)$ and $f(u')$. A device calls the \underline{add\_noise()} function which combines the above $(\varepsilon ,\delta )$-LDP method and D-SGD algorithm to improve the privacy of FL. Meanwhile, the \underline{add\_noise()} function contains two main operations:
\begin{itemize}
\item {Local Update}. Let $D$ be the total number of devices participate in FL. In each iteration, each device calls the \underline{local\_model\_training()} function to compute its local model update. After executing of this function, a device $k$  {obtains} its local models $\{ {\omega^k}\} _{k = 0}^{D}$.
	
\item  {GNM Implementation}. Each device can enforce a certain sensitivity by using scaled versions $\Delta {\bar \omega ^k}$ instead of the true updates, where $\forall k$, $||\Delta {{\bar \omega }^{{k}}}||_2<S$. Dividing the GNM's output by $D$ yields an approximation to the real average of all device  updates, while preventing the leakage of crucial information about an individual, as shown in Fig. \ref{fig-3}.
\end{itemize}
Each device executes the \underline{submit\_model\_update()} function to submit its model update with both the noise-adding gradient information and the payee address for the task publisher.

\textbf{Step 3:} \textit{Evaluation.}
The central aggregator can evaluate a single local model by the following  {operations}.
\begin{itemize}
\item The central aggregator calls the \underline{disclosure\_test\_data()} function that applies the testing dataset for local model evaluation. If the central aggregator fails to  calls the \underline{disclosure\_test\_data()} function, the central aggregator uses its training dataset to evaluate the local model.
\end{itemize}

{At this point, we }note that once the evaluation begins to perform in the central aggregator, it no longer accepts model updates from devices. {The central aggregator calls the \underline{evaluate\_local\_model()} function to select {the qualified devices whose model updates meet the evaluation criteria}. {It supports }the central aggregator to observe the model aggregation results of each round before deciding whether to stop training, thereby reducing the consumption of computing resources and gas.} {Next}, the central aggregator calls the \underline{aggregate\_model\_update()} function to aggregate all local model parameters in the qualified devices to update the current global model. A new global model ${\omega _{t + 1}}$ is allocated by adding this approximation to the current glocal model ${\omega _{t}}$ (see Fig. \ref{fig-2}). Finally, the central aggregator calls the \underline{evaluate\_model()} function to evaluate the updated global model and send the results to the task publisher to complete the current iteration.

\subsection{Update Phase}
After the convergence of the global model update or {the number of iterations} reaches the upper limit, the qualified device calls the \underline{finalize\_contract()} function to {obtain} a reward from the Ethereum blockchain. The central aggregator calls the \underline{post\_evaluation()} function post the results of this FL task. Finally, the task publisher obtains a new better global model.

\section{Solutions to potential issues}\label{models}
Trust and fairness are critical to any marketplaces, {which make it paramount }to design a FL framework where no user can cheat or gain an advantage over other users. Therefore, we further present the following solutions to potential issues related to the proposed framework.

\subsection{Overfitting Problems}
Overfitting is one of the most common problems in machine learning. It reduces the generalization ability of machine learning models. {Our proposed framework is not immune since overfitting problems could occur when } (i) the local training data contains additional noises; (ii) the local training data is insufficient; (iii) the local model is over-trained; (iv) the task publisher releases the testing dataset in advance.

For problems (i), (ii), (iii), the central aggregator can call the \underline{evaluate\_local\_model()} function based on the testing dataset  to solve. For problem (iv), the task publisher and the central aggregator should keep the testing dataset unreleased during the model update phase.

\subsection{Gas Costs Problems}
Participants are compensated for the costs of executing transactions in the blockchain. These costs are called ``gas'' \cite{ref-14} in the Ethereum blockchain. {To avoid missing the mining block tasks, participants set the limit of gas themselves.} Therefore, the upper limit of the gas in SFL framework will result in many devices missing the evaluation of the central aggregator. It means that the smart contracts would never be finalized \cite{ref-5}.

To solve this problem, we perform the following two steps: (i) We allow participants to call evaluation functions locally  {for evaluating} their models. This approach reduces the computation required to evaluate the model, thus avoiding the problem of insufficient gas costs. (ii) The accuracy of the model needed for the device to receive rewards is announced in real-time. Therefore, the device can decide whether to submit a model update based on its local model evaluation result.
\section{PERFORMANCE EVALUATION AND ANALYSIS}

\subsection{Experiment Setting}
To evaluate the performance of the proposed SFL framework, we use Google Tensorflow  {for} the simulation experiments on MNIST dataset and CIFAR-10 dataset, respectively.
Without loss of generality, we assume that the task publisher releases the classification tasks on MNIST and CIFAR-10, respectively. For the two tasks, the participants work together to train the FL model, i.e., AlexNet. \footnote{\href{http://papers.nips.cc/paper/4824-imagenet-classification-with-deep-convolutional-neural-networks}{http://papers.nips.cc/paper/4824-imagenet-classification-with-deep-convolutional-neural-networks}}

{The size of participant set ${P} \in \{10, 100, 200, 300\}$. It is assumed that  ${P}$ contains 20\% malicious participants that launched a poisoning attack, 20\% unreliable participants with low-quality data, and 60\% well-behaved participants.} The training set of well-behaved participants are randomly selected by a uniform distribution of 10 categories. It is assumed that malicious participants mislead the training model by modifying the labels of the training samples to carry out a poisoning attack. We use $\lambda$ as the percentage of the modified labels {that indicate the strength of poisoning attack}, where $\lambda$  in the experiments is set as $5\%, 10\%, 15\%, 20\%$, respectively. The Earth Mover's Distance (EMD) represents the probability distance between a participant's data distribution and the actual distribution for the whole population, which is used as a metric to measure training data quality of the unreliable participants \cite{ref-12}.

We deploy the proposed SFL framework on the Ethereum blockchain. Further, each participant applies D-SGD algorithm to update model parameters iteratively. In the experiments, mini-batch size $m=128$,  learning rate $\alpha = 0.0001$, and  dropout rate $=0.5$. {Like in} \cite{ref-13-13}, we adopt a privacy budget of $\varepsilon =8$, $\delta = \{e^{-1}, e^{-3}, e^{-5}, e^{-6}\}$  to apply the $(\varepsilon ,\delta )$-LDP method, where $\delta$ is the acceptable probability when the differential privacy is unbroken \cite{ref-17}.

\subsection{Performance Analysis}
\begin{figure}[!hptb]
	\centering
	\subfigure []{\includegraphics[width=1\linewidth]{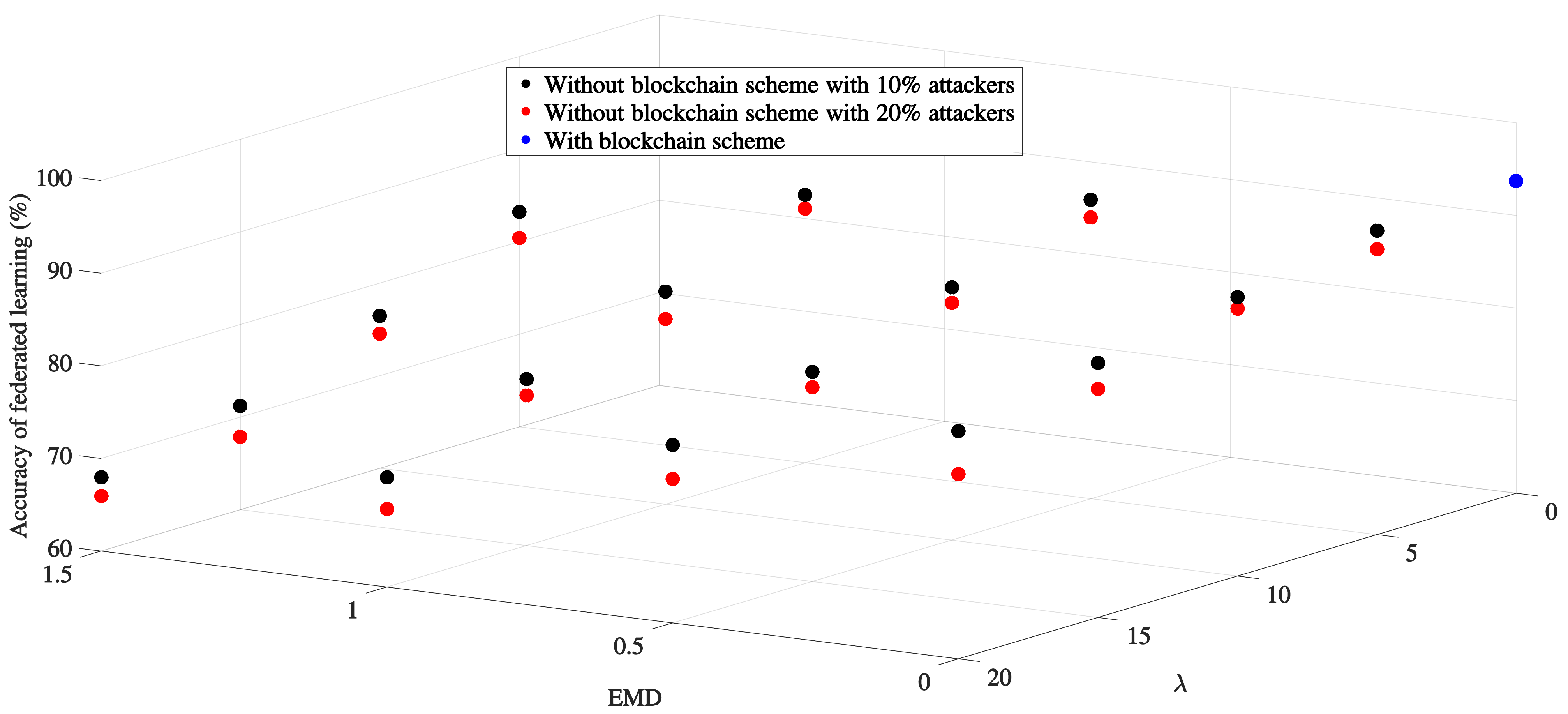}
		\label{a}}
	\hfill
	\subfigure[]{	\includegraphics[width=1\linewidth]{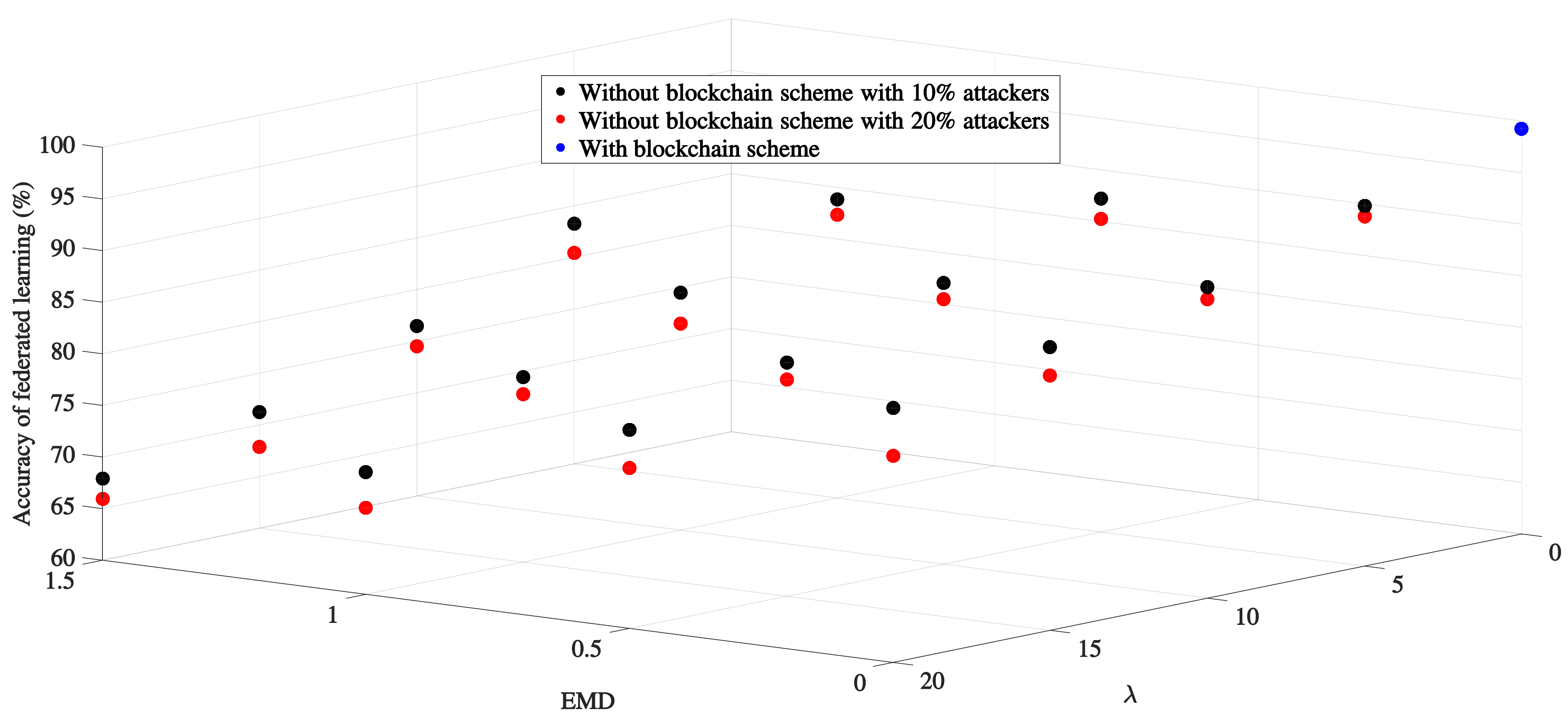}
		\label{b}}
	\caption{The accuracy performances with respect to $\lambda$ and EMD on the datasets: (a) MNIST; (b) CIFAR-10.}
	\label{fig-4}
\end{figure}

{Fig. \ref{fig-4} illustrates the accuracy performances of AlexNet model with different $\lambda$ and EMDs on MNIST dataset and CIFAR-10 dataset, respectively, when $P=100$. It means that the three factors $\lambda$, EMD, and the number of attackers {influence} the accuracy performances. The increase {in} the accuracy of the model is accompanied by {a} decrease in any one of the above factors. {The} results show that the proposed framework can {deter} poisoning attacks. Specifically, as the attack intensity increases, the accuracy of the proposed framework  {remains unchanged}. As shown in Fig. \ref{b}, for example, the learning accuracy is  65.2\% for CIFAR-10 classification task  {with} 20\% malicious participants {exhibiting} the strongest poisoning attack on the AlexNet model (i.e., $\lambda=20\%$, EMD $=1.5$). The central aggregator without defensive measures aggregates a large number of low-quality model updates as the attack intensity increases, thereby reducing the accuracy of the model. It implies that AlexNet model in FL fails to perform image classification tasks under the poisoning attacks.}

\begin{figure}[!t]
	\centering
	\subfigure []{\includegraphics[width=1\linewidth]{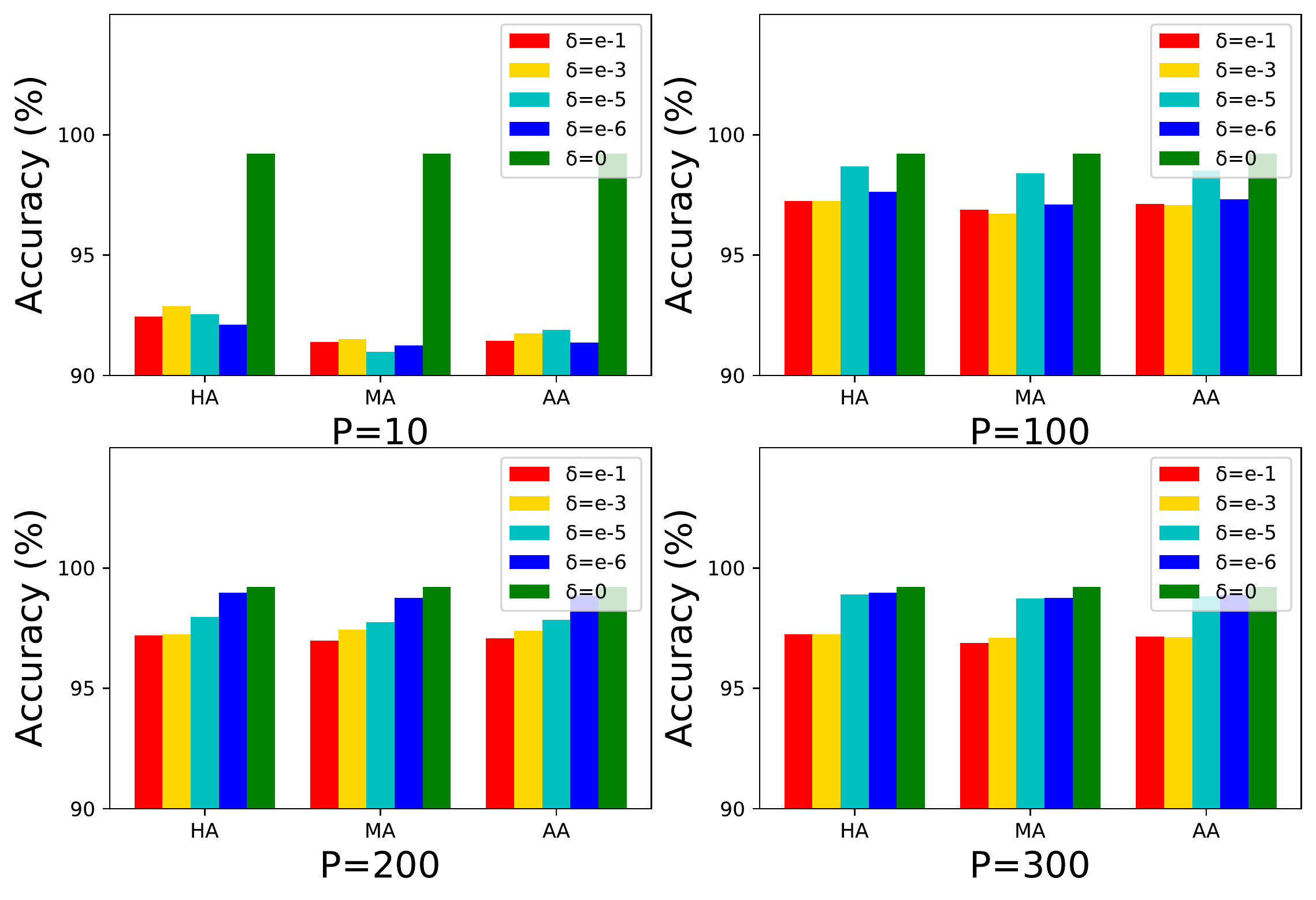}
		\label{a-1}}
	\hfill
	\subfigure[]{	\includegraphics[width=1\linewidth]{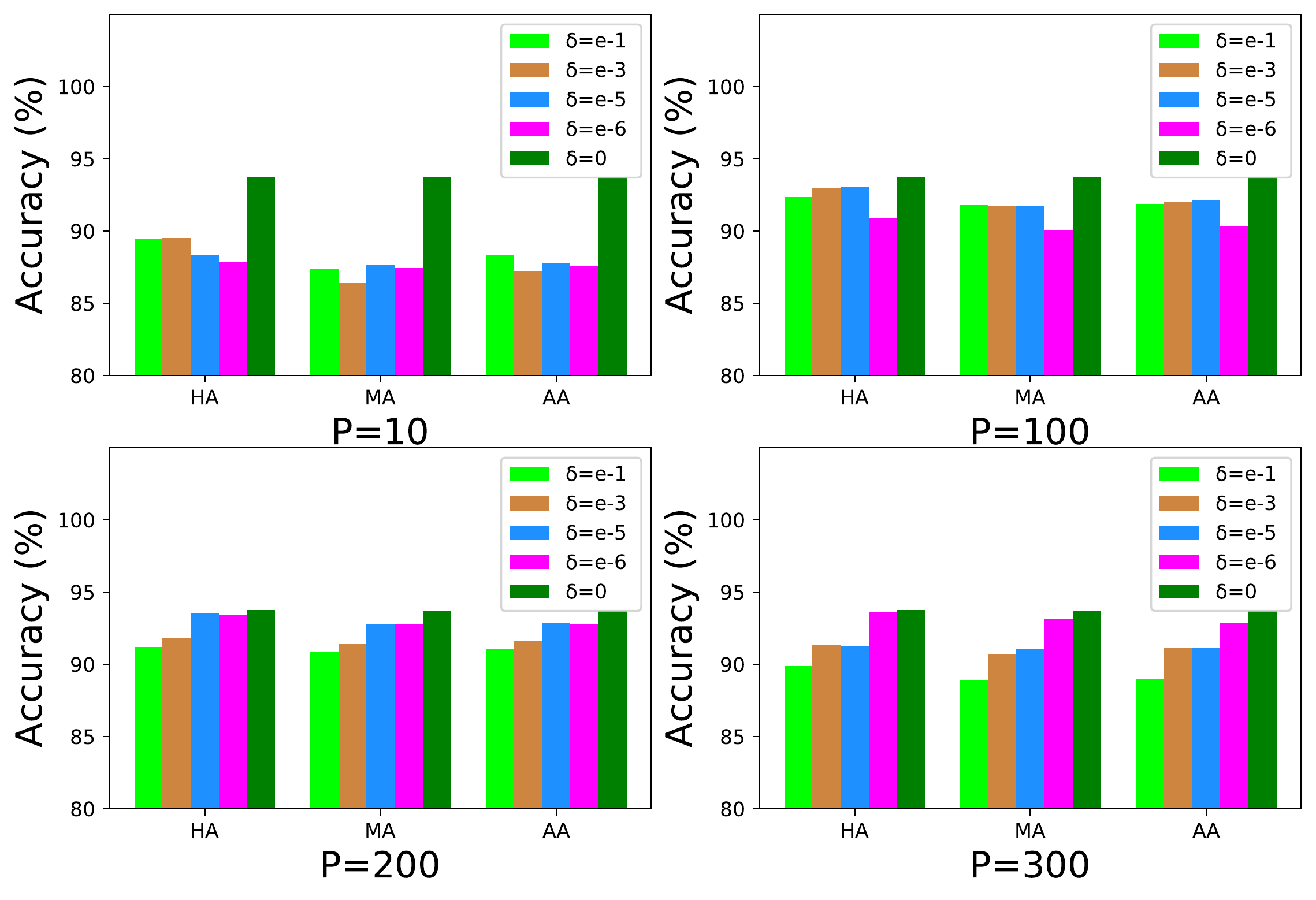}
		\label{b-1}}
	\caption{The accuracy performances with respect to ${P}$, $\varepsilon$, and $\delta$ on the datasets: (a) MNIST; (b) CIFAR-10 }
	\label{fig-5}
\end{figure}

Meanwhile, the well-behaved participants call the \underline{add\_noise()} function to add noises to the gradient information, resulting in the loss of the model accuracy. So, these well-behaved participants may be misidentified as malicious participants because they degrade the accuracy performance of the model. To avoid the confusion of these well-behaved participants, {there is a need to  distinguish them from} malicious participants in the proposed SFL framework. Fig. \ref{fig-5} presents the highest accuracy (HA), minimum accuracy (MA), and average accuracy (AA) by 10 simulations on MNIST dataset and CIFAR-10 dataset, respectively.
We adopt {AA} as the evaluation criteria to determine malicious participants in a FL task. For example, as presented in Fig. \ref{a-1}, when ${P}=100$, the lowest $AA$ is obtained as 90.32\% that is the same as the accuracy of model in Fig. \ref{a} with $\lambda=5$ and EMD $=1.5$. Here,  $AA$ (i.e., $\lambda=5$, EMD $=1.5$) can be used as the threshold to determine malicious participants. The central aggregator calls the \underline{evaluate\_local\_model()} function on the testing dataset to evaluate the local model uploaded by the participants. If the result of the above function is lower than $AA$,  a reward will not be paid to the participant who may be a malicious participant. If not, the participant calls the \underline{finalize\_contract()} function to get a reward.

{{For Fig. \ref{fig-5}}, we compare the accuracy of the proposed framework with that of a typical FL framework (i.e., non-differentially private (Non-DP) model) at different defense strengths of inference attacks. Numerical results show a trade-off between the protection  {from} inference attacks and the accuracy of the model. Specifically, the accuracies of Non-DP model are 99.21\% and 93.72\% on MNIST and CIFAR-10 datasets, respectively. When $\delta=e^{-6}$, $P=300$, Fig. \ref{fig-5} illustrates that the accuracy of the proposed framework is close to that of the scheme without defensive measures, which indicates  that the local differential privacy mechanism prevents inference attacks while sacrificing some accuracy. Besides, the central aggregator uses a batch average aggregation mechanism (see Fig. \ref{fig-2}) to {mitigate} the influence of noises on the accuracy of the model. Therefore, within a threshold of a certain number of participants (i.e., in this study, the threshold is set as 300), the more participants involved in the FL, the better performance  achieved.}

\section{CONCLUSIONS AND FUTURE DIRECTIONS}
In this article, we addressed data privacy leakage issues {related to ensuring} secure FL in 5G networks. We proposed a blockchain-based framework to defend against poisoning attacks. In the proposed framework, a market is created to trade model updates based on smart contracts in blockchain to validate the model updates against poisoning attacks automatically. Additionally, we introduce the local differential privacy technique in smart contracts to prevent the membership inference attacks. Specifically, we add well-designed Gaussian noises to the model updates uploaded by the participants to defend against membership inference attacks. We validate our proposed security framework on two datasets, which yielded secure FL to 5G networks. {{The study has several open questions that need to be further investigated}.
First, because the federated averaging algorithm is used to aggregate the model updates, we encounter unfair usage to certain devices. Second, efficiency issues arise since the use of deep learning algorithms increases the computational demands of the proposed system. Third, the more participants in the proposed framework, the greater the communication overhead, which in turn reduces the accuracy of FL. These are among the many issues that need to be addressed in the road to fair and efficient blockchain-based FL frameworks of the future.}

\section*{Acknowledgment}
This work is supported in part by the Science and Technology Project of State Grid Corporation of China (Grant No. SGHL0000DKJS1900883), in part by AMI acknowledges funding from the Prince Sattam Bin Abdulaziz University, Saudi Arabia via the Deanship for Scientific Research funding for the Advanced Computational Intelligence \& Intelligent Systems Engineering (ACIISE) Research Group Project Number 2019/01/9862, in part by the National Research Foundation (NRF), Singapore, Singapore Energy Market Authority (EMA), Energy Resilience, NRF2017EWT-EP003-041, Singapore NRF2015-NRF-ISF001-2277, Singapore NRF National Satellite of Excellence, Design Science and Technology for Secure Critical Infrastructure NSoE DeST-SCI2019-0007, A*STAR-NTU-SUTD Joint Research Grant on Artificial Intelligence for the Future of Manufacturing RGANS1906, WASP/NTU M4082187 (4080), Singapore MOE Tier 2 MOE2014-T2-2-015 ARC4/15, and MOE Tier 1 2017-T1-002-007 RG122/17.
\ifCLASSOPTIONcaptionsoff
  \newpage
\fi

\bibliographystyle{IEEEtran}
\bibliography{reference}
\begin{IEEEbiographynophoto}{YI LIU}
    [S'19] (97liuyi@ieee.org) received the B.Eng. degree in network engineering from Heilongjiang University, Harbin, China, in 2019. He is currently pursuing a Ph.D. degree at the Faculty of Information Technology, Monash University, Melbourne, Australia. His research interests include security \& privacy, federated learning, edge computing, and blockchain.
\end{IEEEbiographynophoto}
\begin{IEEEbiographynophoto}{JIALIANG PENG}
	(jialiangpeng@hlju.edu.cn) received the B.S. and M.S. degrees in Computer Science from Heilongjiang University, Harbin, China, and the Ph.D. degree in Computer Science at Harbin Institute of Technology, Harbin, China. He is an associate professor in the School of Data Science and Technology, Heilongjiang University, China. He worked as a Post-Doctoral Scholar in Norwegian University of Science and Technology, Gjøvik, Norway. His research interests include biometric recognition, data management, and cyber-space security.
\end{IEEEbiographynophoto}
\begin{IEEEbiographynophoto}{JIAWEN KANG}
	(kavinkang@ntu.edu.sg) received the M.S. degree from the Guangdong University of Technology, China, in 2015, and the Ph.D. degree at the same school in 2018. He is currently a postdoc at Nanyang Technological University, Singapore. His research interests mainly focus on blockchain, security and privacy protection in wireless communications and networking.
\end{IEEEbiographynophoto}
\begin{IEEEbiographynophoto}{ABDULLAH M. ILIYASU}
	(abdul-m-elias@ieee.org) received the M.E. and Dr.Eng. degrees in computational intelligence and intelligent systems engineering from the Tokyo Institute of Technology (Tokyo Tech.), Japan. His research interests mainly focus on computational intelligence, quantum cybernetics, quantum image processing, quantum machine learning, cyber and information security, hybrid intelligent systems, the Internet of Things, 4IR, health informatics, and electronics systems reliability.
\end{IEEEbiographynophoto}
\begin{IEEEbiographynophoto} {DUSIT NIYATO} 
	[M'09, SM'15, F'17] (dniyato@ntu.edu.sg) is currently a professor in the School of Computer Science and Engineering, Nanyang Technological University. He received his B.Eng. from King Mongkut's Institute of Technology Ladkrabang, Thailand, in 1999 and his Ph.D. in electrical and computer engineering from the University of Manitoba, Canada, in 2008. His research interests are in the area of energy harvesting for wireless communication, the Internet of Things, and sensor networks.
\end{IEEEbiographynophoto}
\begin{IEEEbiographynophoto} {AHMED A. ABD EL-LATIF} 
 	(aabdellatif@nu.edu.eg) received the B.Sc. degree with honor rank in Mathematics and Computer Science and M.Sc. degree in Computer Science from Menoufia University, Egypt. He received Ph.D. degree in Computer Science at Harbin Institute of Technology, Harbin, China. He is an associate professor of Computer Science at Menoufia University, Egypt and School of Information Technology and Computer Science, Nile University, Egypt. He is a fellow at Academy of Scientific Research and Technology, Egypt. His areas of interests are multimedia content encryption, secure wireless communication, IoT, applied cryptanalysis, perceptual cryptography, secret media sharing, information hiding, biometrics, forensic analysis in digital images, and quantum information processing.
\end{IEEEbiographynophoto}

\end{document}